\documentstyle[11pt,newpasp,twoside,psfig]{article}
\def\edcomment#1{\iffalse\marginpar{\raggedright\sl#1\/}\else\relax\fi}
\reversemarginpar

\begin{document}
\title{The Recurrent Nova U~Scorpii -- A Type Ia Supernova Progenitor} 

\author{T. D. Thoroughgood, V. S. Dhillon, S. P. Littlefair} 

\affil{Department of Physics \& Astronomy, University of Sheffield, Sheffield 
S3 7RH, UK}

\author{T. R. Marsh} 

\affil{Department of Physics \& Astronomy, University of Southampton, 
Southampton SO17 1BJ, UK}

\author{D. A. Smith} 

\affil{Winchester College, Winchester SO23 9LX, UK}

\begin{abstract}

We derive the mass of the white dwarf in the eclipsing recurrent nova U~Sco 
from the radial velocity semi--amplitudes of the primary and secondary stars. 
Our results give a high white dwarf mass of 
$M_1 = 1.55 \pm 0.24M_\odot$, consistent with the thermonuclear runaway 
model of recurrent nova outbursts. We confirm that U~Sco is the best Type 
Ia supernova progenitor known, and predict that the time to explosion is 
within $\sim 700\,000$ years.

\end{abstract}

\noindent{This work has been accepted for publication in the {\em Monthly Notices of the 
Royal Astronomical Society} (astro-ph/0107477).}

\section{Recurrent Novae as Type Ia Supernova Progenitors}

Type Ia supernovae (SNe Ia) are believed to represent 
the thermonuclear disruptions of C--O white dwarfs, when these 
white dwarfs reach the 
Chandrasekhar limit and ignite carbon at their centres. The 
favoured scenario for SN Ia progenitors is the single--degenerate model, 
in which the WD accretes from a subgiant or giant companion (Livio 2000). 

Suspected SN Ia progenitors are the recurrent novae (RNe) -- 
a small class of cataclysmic variables (CVs) which show repeated nova 
outbursts on a timescale of decades. To produce such short intervals between 
outbursts, the thermonuclear runaway model requires a near--Chandrasekhar 
mass white dwarf (Starrfield, Sparks \& Shaviv 1988).

Despite the ejection of material during a nova event, most of 
the accreted material remains on the surface of the white dwarf (Hachisu et 
al. 2000). Further accretion of mass onto a massive white dwarf will result 
in the Chandrasekhar limit being overcome, and the subsequent detonation of 
the white dwarf to produce a SN Ia. An observational 
determination of the white dwarf mass in RNe is therefore crucial in 
assessing whether RNe are SN Ia progenitors.

\section{The Mass of the White Dwarf in the Recurrent Nova U~Sco}

We present spectroscopy of the eclipsing recurrent nova U~Sco. The radial 
velocity semi-amplitude of the primary star was found to be $K_W = 
93 \pm 10$ km\,s$^{-1}$ from the motion of the wings of the HeII$\:
\lambda4686$\AA$\:$ emission line (Fig.1). By detecting weak absorption 
features 
from the secondary star, we find its radial velocity semi-amplitude to be 
$K_R = 170 \pm 10$ km\,s$^{-1}$ (Fig.2). From these parameters, we obtain a 
mass of 
$M_1 = 1.55 \pm 0.24M_\odot$ for the white dwarf primary star and a mass of 
$M_2 = 0.88 \pm 0.17M_\odot$ for the secondary star. The radius of the 
secondary is calculated to be $R_2 = 2.1\pm0.2R_\odot$, confirming that 
it is evolved. The inclination of the system is calculated to be 
$i = 82.7^\circ\pm2.9^\circ$, consistent with the deep eclipse seen in the 
lightcurves. The helium emission lines are double-peaked, with the 
blue-shifted regions of the disc being eclipsed prior to the red-shifted 
regions, clearly indicating the presence of an accretion disc. The high mass 
of the white dwarf is consistent with the thermonuclear runaway model of 
recurrent nova outbursts, and confirms that U~Sco is the best Type Ia 
supernova progenitor currently known. 
Assuming an average white dwarf growth rate of $\dot M_{av} 
\sim 1.0 \times 10^{-7} M_\odot$ y$^{-1}$ (Hachisu et al. 2000), we predict 
that U~Sco is likely to explode within $\sim 700\,000$ years.

\begin{figure*}
\begin{tabular}{ll}
\psfig{figure=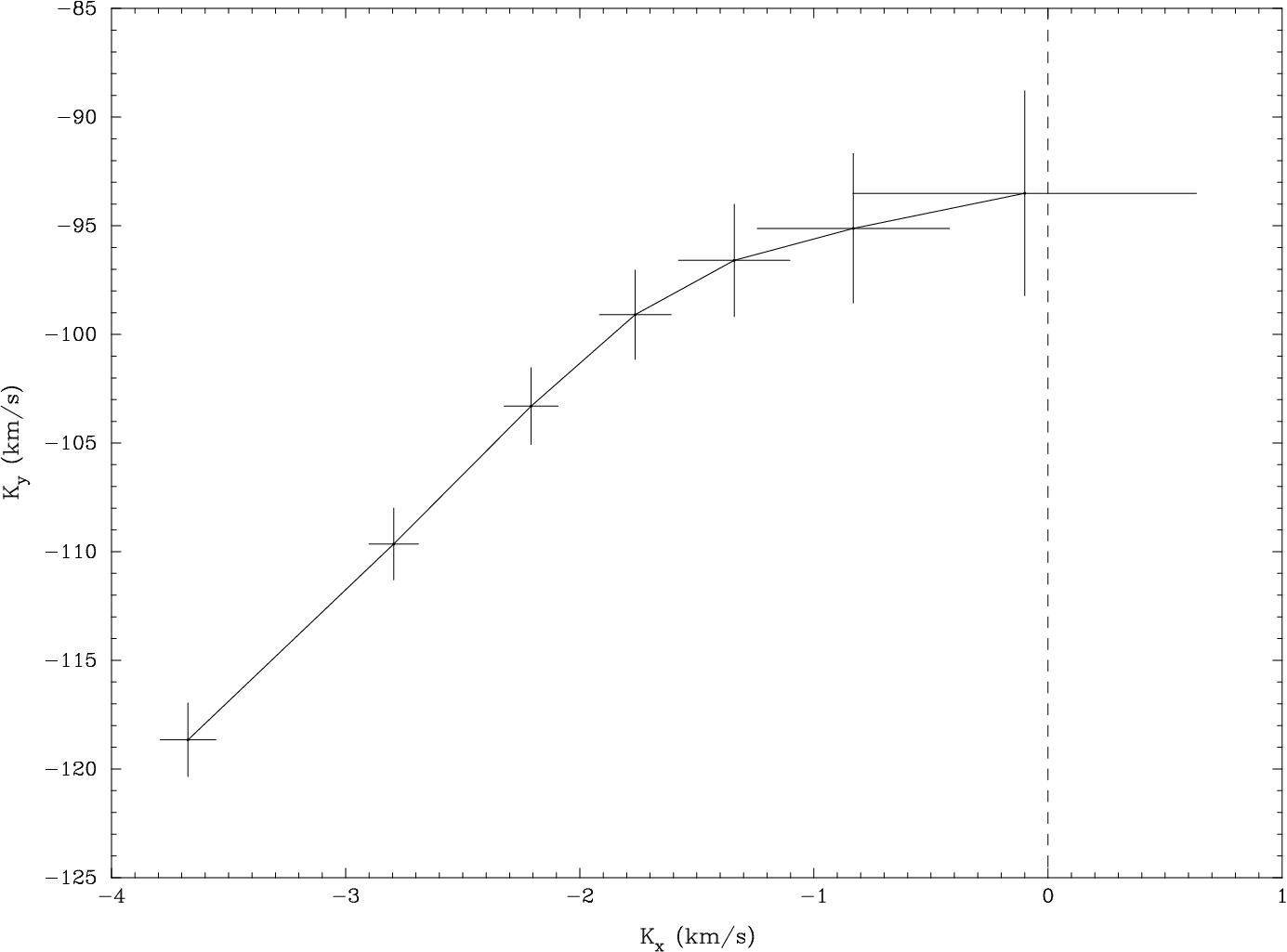,width=6cm,angle=-90} &
\hspace{0.8cm}\psfig{figure=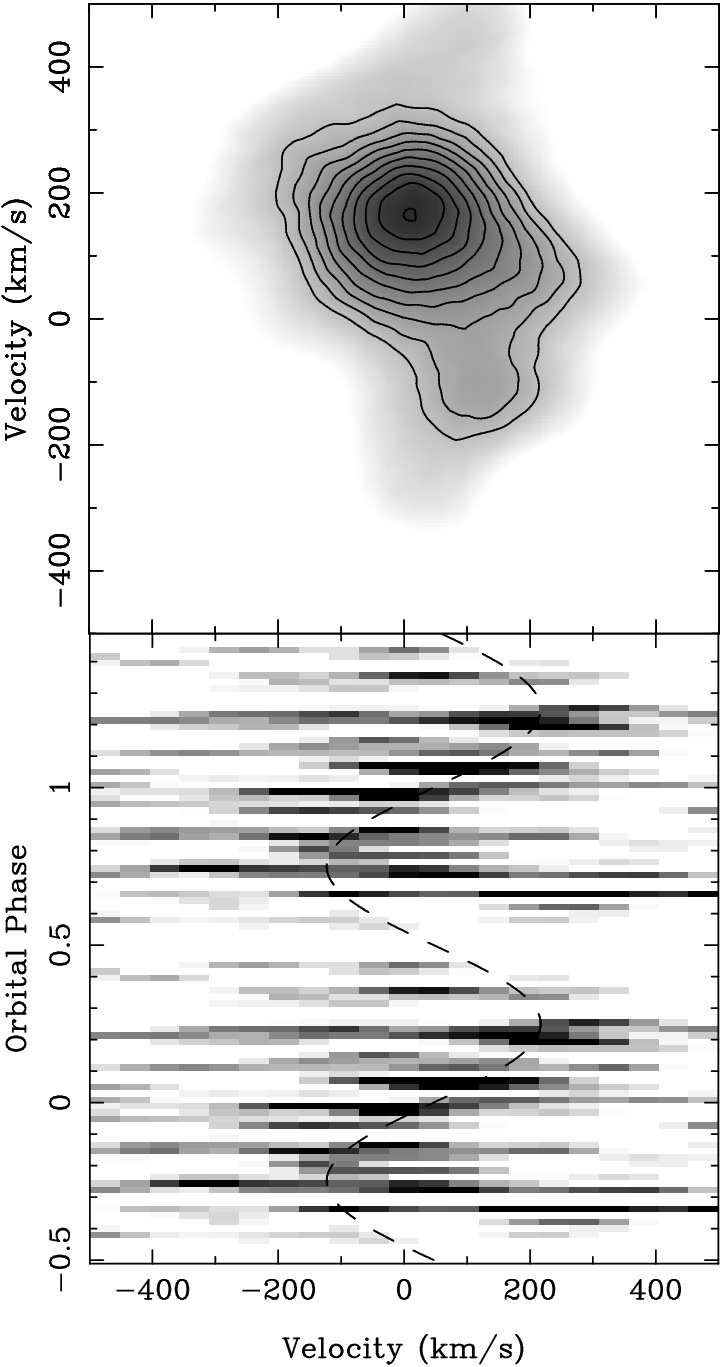,width=3.9cm,angle=0} \\
{\bf Fig.1 --} The light centres diagram for &
{\bf Fig.2 --} The skew map (top), and \\
HeII$\:\lambda4686$\AA$\:$; $K_W = 93\pm10$ km\,s$^{-1}$. &
trailed cross-correlation functions;\\ 
 & dashed line shows  $K_R = 170$ km\,s$^{-1}$
\end{tabular}
\label{fig:radials}
\end{figure*}


\begin{references}

Hachisu, I., Kato, M., Kato, T. \& Matsumoto, K. 2000, ApJ, 528, L97 \\ 
Livio, M., 2000, in ``Supernovae and Gamma Ray Bursts'', STScI Symp.13 CUP \\
Starrfield, S., Sparks, W. M. \& Shaviv, G. 1988, ApJ, 325, L38 \\
Thoroughgood, T. D. et al. 2000, MNRAS, in press (astro-ph/0107477)
\end{references}
\end{document}